# Demonstration of a Response Time Based Remaining Useful Life (RUL) Prediction for Software Systems

Mohammad Rubyet Islam[a, 1], Peter Sandborn[a]

[a] Department of Mechanical Engineering, University of Maryland, College Park, USA

**Abstract**

Prognostic and Health Management (PHM) has been widely applied to hardware systems in the electronics and non-electronics domains but has not been explored for software. While software does not decay over time, it can degrade over release cycles. Software health management is confined to diagnostic assessments that identify problems, whereas prognostic assessment potentially indicates when in the future a problem will become detrimental. Relevant research areas such as software defect prediction, software reliability prediction, predictive maintenance of software, software degradation, and software performance prediction, exist, but all of these represent diagnostic models built upon historical data – none of which can predict an RUL for software. This paper addresses the application of PHM concepts to software systems for fault predictions and RUL estimation. Specifically, this paper addresses how PHM can be used to make decisions for software systems such as version update/upgrade, module changes, system re-engineering, rejuvenation, maintenance scheduling, budgeting, and total abandonment. This paper presents a method to prognostically and continuously predict the RUL of a software system based on usage parameters (e.g., the numbers and categories of releases) and performance parameters (e.g., response time). The model developed has been validated by comparing actual data, with the results that were generated by predictive models. Statistical validation (regression validation, and k-fold cross validation) has also been carried out. A case study, based on publicly available data for the Bugzilla application is presented. This case study demonstrates that PHM concepts can be applied to software systems and RUL can be calculated to make system management decisions.

**Keywords:** Prognostic and Health Management (PHM), Software, Performance and Usage Parameters, Optimization, Remaining Useful Life (RUL), Natural Language Processing (NLP), User Interface (UI), User Experience (UX), Combined Predictive Variable (CPV).

---

[1] Corresponding author. E-mail address: rubyet@umd.edu

## 1. Introduction

System failures result in the need for expensive and time-consuming maintenance actions. Systems fail due to a myriad of causes that are linked to more than just their hardware components. System failures due to software result in significant economic impacts on a large number of people, Table 1.

**Table 1**. Typical Cost of Software Failures in 2018 [1]

| Failures Studied | 606 failures from 314 companies |
|---|---|
| Financial Loss | US $1.7 trillion |
| People Affected | 3.6 billion |
| Lost to Downtime | 268 years |

Prognostics and health management (PHM) methods reduce the time and cost for the maintenance of products or processes through efficient and cost-effective diagnostic and prognostic activities. PHM systems use real-time and historic state information of subsystems and components to provide actionable information, enabling intelligent decision-making for improved performance, safety, reliability, and maintainability. The goal of PHM is to provide actionable information to assist in system support decision making. While diagnostics is the process of detecting and isolating faults or failures, prognostics is the process of predicting the future state or remaining useful life (RUL) based on current and/or historic conditions [2]. Prognostics is based upon the understanding that systems/products fail after a period of degradation, which, if measured (via an RUL), can be used to prevent system breakdown, avoid collateral damage, and minimize operation costs [3]. Data on degradation parameters is used to regress selected mathematical model(s) and to extrapolate the model(s) to a pre-determined failure threshold. The difference between the time (or other appropriate usage parameters) at the current point and the point at which the extrapolated behavior exceeds the failure threshold is the estimated RUL [4]. Prognostics is focused on predicting the time (or other appropriate usage parameters) at which a system or a component will no longer perform its intended function. The inability of the system to perform its intended function is most often a failure beyond which the system can no longer be used. The predicted time-to-failure is referred to as the remaining useful life (RUL) [5]. Remaining useful life (RUL) represents the useful life left in an asset at a particular point in its operation. The estimation of RUL is central to condition-based maintenance and prognostics and health management. However, diagnostics are vital for successful prognostics because an acceptable prognostic method starts with robust diagnostics [6,7]. Approaches that are used to build prognostic models can be categorized broadly into data-driven, model-based, and fusion[2] [8]. The solution proposed in this paper for software systems is a fusion approach where both data-driven and model-based approaches have been utilized.

---

[2] Fusion prognostics methods fuse data-driven methods and model-driven methods to predict the remaining useful life of products.

A significant body of research exists on software defect prediction, reliability prediction, predictive maintenance, usability, and other similar topics. However, no research exists on the application of Prognostic and Health Management (PHM) or Remaining Useful Life (RUL) to software systems. This research focuses on the application of PHM to faults/failures and enhancement requests due to version update/upgrade of individual software components that are integrated to create a software system. Also, all the relevant predictive models, that have been developed in the past for software, are diagnostic in nature (not prognostic). This paper applies *dynamic* PHM concepts to software systems and calculates the RUL using inputs from embedded sensors/health monitoring tools and reports from the end users and the development team[3].

## 1.1. Software Degradation and Failure

Although, no known research exists on the application of PHM to software systems, relevant research is available on the diagnostic health management of software systems. In Section 2 we will discuss how the application of PHM to software systems lacks the calculation (or even the definition) of remaining useful life (RUL). The application of PHM to software systems requires identification of appropriate target and predictive parameters and usage parameters.

Technically, software does not decay over time but can degrade over release cycles. Software degradation, also known as: software decay, aging, rotting, smell, degradation, code rot, bit rot, software entropy, and software erosion, represents a slow deterioration of software performance over time or its diminishing responsiveness that will eventually lead to software becoming faulty, unusable, and in need of update/upgrade. Software degradation is a common problem faced by legacy systems (a legacy system is a system that is based on methodologies, processes, architectures, technologies, parts, and/or software that is out-of-date, i.e., old). Software degradation creates the need for updates to accommodate the changing environment in which the software resides.

Software decay can happen because of unused or leftover code that may contain bugs that were not resolved during changes made in the code base, environments, and components. Other reasons include the lack of software updates/upgrades, improper maintenance, incompatible changes in architecture, inappropriate integration, memory leaks, data corruptions, and unused code. Since PHM can prognostically measure decay/degradation over time and since software decays/degrades due to changes over releases, with the use of appropriate parameters and definitions, PHM should be applicable to software systems. However, unlike hardware, software errors do not develop over time. Rather, they are introduced over releases as flaws and errors at every stage of the software life cycle. Hence, *release cycle* is an appropriate *usage* parameter. The parameters that govern both software releases and software performance are the faults and enhancements. Software update, sometimes called a software patch, is a download for an application, operating system, or software suite that provides fixes for features that are not working as intended or adds minor software enhancements and compatibilities. Whereas a software upgrade is a new version of software that offers a significant change or *major* improvements over the current version.

---

[3] End users are the people that a software program is designed for. Whereas, the software development team consists of developers, software quality assurance engineers, user experience (UX) designers, business analysts and product owners (project managers).

Software failure corresponds to unexpected runtime behavior, observed by the end users. Whereas a fault is a static software characteristic that causes a failure to occur. Faults are the result of incorrect design choices, simple human blunders, or the forces of nature [9]. The terms bug and defect are often used to describe fault, error, or even failure, but are not precise enough [10]. In this paper we used the term 'fault' and 'failure' instead of bug/defect to avoid confusion. Prediction of the RUL of software can save money by reducing unexpected failures and by assisting in planning preventive maintenances and releases.

## 1.2. Paper Objective and Organization

The goal of this paper is to introduce dynamic PHM models to software systems to estimate the software system's RUL. Our target is to predict a software system's health condition over release cycles and to identify the point (before the software fails) at which a decision needs to be made for a software system to be updated/upgraded, rejuvenated, re-engineered or abandoned. Existing PHM techniques are used in this paper to build a fusion solution, which is not in and of itself new, but it's application to a software system is new and has not been previously demonstrated.

This paper is organized as follows. Section 2 presents background on existing research that has been carried out in related software domains. This section also discusses the differences between PHM in hardware and in software. In Section 3, a methodology is proposed to prognostically estimate the RUL of software using a fusion approach. Section 4 presents a demonstration of the method using an open-source software system. Conclusions and future research directions are summarized in Section 5.

## 2. Literature Review

This section reviews the relevant literature associated with software reliability prediction, software predictive maintenance, software decay, software performance prediction and software defect prediction methods. We will also distinguish software reliability from software PHM.

**2.1 Software Reliability**

According to ISO9126, software reliability is the capability of a software product to maintain a specified level of performance, when used under specified conditions for a stated period [11-14]. Predictions of reliability help to avoid failures and large disturbances [7,15] and increase the reliability of systems. Software reliability is focused on the probability of failure free operation and the primary goal is to identify the probability of a software system failing in a given time interval (mean time between failure) due to faults [5]. However, software does not only fail due to faults but can also fail due to the implementation of future enhancements, and not all enhancements necessarily generate faults. As explained in Section 1.1, software does not decay. over time, but performance fluctuates as enhancement requests are implemented and faults are generated over future releases. Future enhancement requests, as a source of software failure, are not accounted for in traditional software reliability analysis. PHM, prognostically, considers the impacts of both the faults and future enhancement requests. Software reliability is calculated as a function of time whereas the proposed software RUL is calculated as a function of release cycles. Unlike reliability, PHM is not limited to any time boundary and is dynamic. The important attributes for measuring software reliability include Mean Time to Failure (MTTF), Mean Time to Repair (MTTR), Mean Time Between Failure (MTBF), Probability of Failure on Demand (POFOD), etc. [16]. Software reliability is, mainly, calculated using static methods with historical (diagnostic) data that was collected in a designated time frame. Alternatively, PHM estimates RUL based on both diagnostic and prognostic data that includes expected future enhancements to the software.

Even though multi-stage reliability models in the software domain exist, reliability does not estimate RUL. Software reliability growth models assume that reliability growth is due to bugs being removed. The RUL of software is impacted by future enhancement requests or the addition of new features that may not generate any faults[4]. Reliability is a measurement of performance attributed to faults and the RUL of software is attributed to performance that may or may not be attributed to faults. Reliability is measured using failures whereas PHM addresses the combination of both failures and future enhancements. Most commonly, reliability prediction is carried out in the design stage. Whereas PHM and RUL estimation is carried out continuously at the production and maintenance stages.

---

[4] As an example, upgrading the version of a load balancer that improves the distribution of software application traffic may not populate any errors, but will improve software performance, thereby increasing the software's RUL. Alternatively, the implementation of new graphics may not generate any faults but could reduce performance, thereby decreasing the software's RUL.

**2.2 Other Relevant Software Literature**

Software maintenance is the process of modifying a software system or component after delivery to correct faults, improve performance or other attributes, or adapt to a changed environment [17]. PHM can be used as a tool to plan predictive maintenance activities.

Software Defect Prediction (SDP) predicts defects in modules that are defect prone and require extensive testing [18, 19]. Defect prediction is, also, mainly a diagnostic process, based on historical data. Defect prediction does not account for future enhancement requests and does not predict remaining useful life (RUL). PHM takes data inputs dynamically that are generated by sensors in the health monitoring tools and by the end users and the development team, in the production environment. Defect prediction is commonly performed during software development phases and less common in the production environment. Defect prediction methods can potentially be used as elements for RUL estimations.

Integrated Software Health Management (ISWHM) [20] is a health monitoring system that monitors on-board software and sensors to detect failures both in hardware and in software components using probabilistic modeling like Bayesian networks. ISWHM is a diagnostic approach that does not address future enhancement requests. However, methods used in ISWHM to collect fault data can be utilized in data collection for PHM. Models to evaluate software decay [21, 22] are used to understand software degradation mechanisms.

Albu and Popentiu-Vladicescu (2013) [23] attempted to predict response time for web services and web applications. They measure response time/execution time at a specified point in time based on the existing system, however, they did not consider the influences of faults and future enhancement requests on performance parameters.

While there has been work done on finding latent faults that have been resident in software since its original development, much less works exist on finding problems that are introduced after the software is deployed to the production environment. Once software is fielded, faults are generally reported only when a problem arises instead of being continuously assessed. As of today, all of the suggested and practiced solutions for software are diagnostics in nature, not prognostic.

**3. Proposed Solution**

The research described in this paper dynamically utilizes fault, enhancement, and performance data that are associated with software systems. The proposed architecture shown in Figure 1 includes three modules: *Data Acquisition and Pre-processing*, *Optimization*, and *RUL Prognosis*. In the first step, target variables and predictive variables are identified. Target variables are the performance parameters (see Section 3.1), and the predictive variables are the faults and future enhancement requests that are related to the corresponding target variables. Data on the variables are collected using software health monitoring tools, from reports populated by the end users and the development team. The collected data is pre-processed (cleaning and sanitizing) using a data quality assessment metrics.

In the optimization phase, pre-processed data are categorized using a custom-built categorization matrix. This matrix classifies faults/enhancement requests into categories as such server, network, database, configuration, and others, and into sub-categories such as overloaded server, sluggish client, slow network services and others. These divisions provide better scalability and ease in mapping the predictive variables to their corresponding target variables. Such mapping, also, helps users understand the severity and the extent of impacts of faults/enhancements on performance parameters. After categorization, an NLP-based classification (Naïve Bayes) is carried out to classify and size faults and enhancement requests. Here, sizing is the process of estimating faults and enhancements. Estimation is carried out based on story points (explained in Section 3.3). Upon estimation, story points are individually multiplied with impact factors and summed to estimate a Consolidated Predictive Variable (*CPV*). The equation used to calculate *CPV* for each software release is:

$$CPV = (\pm SP_1)(IF_1) + (\pm SP_2)(IF_2) + \ldots + (\pm SP_n)(IF_n) \tag{1}$$

where *SP* is the story point (which can be positive or negative, see Section 3.3) and *IF* is the impact factor.

The *CPV* is used to calculate the target variable(s). Clusters of analogous releases are formed using k-means clustering for better fit of the prognostic model(s).

To select appropriate prognostic model(s), factors such as: problem type (supervised/unsupervised), variable types (structured/un-structured/semi-structured), data volume, relationship between variables, number of variables and others are considered. Model(s) selection may vary depending up on software system types, data volume, relationship between variables, target variable types and many other factors. Prognosis of the target variable is carried out based on the *CPV* from analogues releases. Prognostic values are used to estimate the RUL by plotting them across usage parameter(s). Data on predictive variables are continuously collected and used in the proposed model to dynamically update the estimated RUL.

### 3.1 Performance and Usage Parameters

The following is a list of software performance parameters that are widely used in the industry [17, 18, 24-26]:

- **Execution time:** Time spent by the software system executing the task [26]
- **Timeliness:** Timeliness is measured by the software response time and throughput [17]
- **Throughput:** The rate at which the system can process inputs [24-27].
- **Response time (RT):** The time between the end user requesting a response from the application and a complete reply [26].
- **Availability:** The measure of availability is the probability that a system is available for use, considering repairs and other down-time [26].
- **Utilization:** The percentage of the theoretical capacity of a resource that is being used [26].

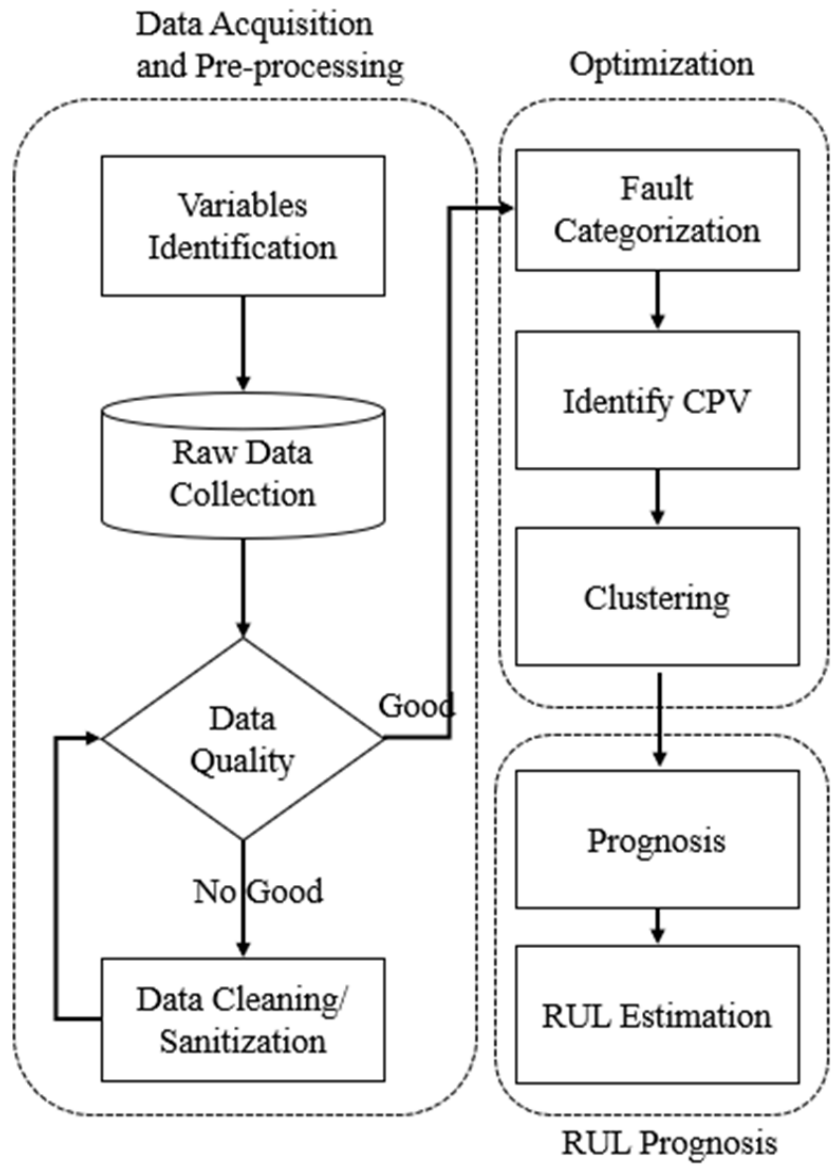

**Fig. 1**. Prognostic model.

Since, unlike hardware, software does not decay over time, time is not an appropriate usage parameter. Software rots/degrades/decays as changes are made, and these changes happen over releases. Hence, *release cycle* is a better usage parameter. As a result, in this methodology, RUL is measured over releases rather than over time.

### 3.2 Data Collection and Pre-processing

After the selection of appropriate target variables and predictive variables, data collection is carried out (both systems generated, and reported by the end users and the development team). While training the model(s), the predictive variables, also known as input variables, are estimated based on *newly emerged* and *resolved* faults and *implemented* expected enhancements. Only the newly emerged and resolved faults and implemented enhancements are responsible for any changes in software performance in the relevant releases. For prognosis, unresolved faults and future enhancement requests are used to architect release contents.

Historical data on the target variables are needed to train the prognostic model(s). To collect this data, individual test beds are prepared for each software release and target variables are measured using appropriate methods and tools. Data is pre-processed by cleaning and sanitizing using quality assessment metrics. The metrics incorporate data quality dimensions (e.g., completeness, accuracy, timeliness, usability), sub-dimensions (e.g., depth and breadth of data record, spelling errors, duplicate values) and quality criteria questions. This ensures that collected data are of quality and mapped with the right target variable(s).

### 3.3 Optimization

Initially, fault and future enhancement request data are classified into multiple categories and sub-categories based on a category metrics. Categorization improves the estimation of story points, impact factors and *CPV* by providing the necessary scaling and grouping. During the formulation of prognostic model(s) and estimation of the RUL, the goal should be to combine several features into an optimal prognostic parameter that can easily be modeled with a chosen function [28]. The rationale behind the fusion of the features is to combine several information sources into one predictive parameter that has improved robustness and an underlying trend related to the overall condition of the software. If the trend is not clear and is not defined over the system's lifetime, the difficulties in analysis can increase, and the accuracy can decrease. Therefore, reducing dimensions by selecting the best feature is necessary to remove the irrelevant and erroneous features. In the second step, each fault and enhancement requests are estimated in story points. Though there are many estimation techniques, story pointing at present, is the most widely used estimation technique [29]. Story points are a measure for expressing the overall size of faults/enhancement requests. Story points consider attributes like volume of work, inherent risks, complexities, and the effort required to complete all faults and enhancement requests. There are many different scales to measure story points, among which Fibonacci sequence is the most widely used. Each number in the Fibonacci scale is exponentially larger (by about 60%) than the previous number which helps users to recognize the differences more easily and to define the complexity of each story point. For this paper the Fibonacci sequence [24] used was 1,2,3,5, and 8. Where, 1 represents the smallest and 8 the largest size of story points. An NLP-based technique has been used as the algorithm to learn and estimate sizes of story points automatically. Larger story points do not necessarily mean higher impacts on the target performance parameter. Hence, an impact scale with appropriate impact factors has been introduced. It is a common practice that impact factors are scaled as critical, major, medium, and minor and it is ideal to convert them with an increment of 0.5. However, this can change depending on the user group. As long as the user uses the same scale and conversion throughout the life cycle of a software system this model will be valid. Table 2 shows the conversion between impact scales and impact factors. The higher the factors, the higher the impacts.

**Table 2.** Impact factor table– Conversion between impact scales and numeric factors.

| Impact Scale | Impact Factor |
|---|---|
| Critical | 1 |
| Major | 0.75 |
| Medium | 0.5 |
| Minor | 0.25 |

The *CPV* of individual faults/enhancement requests are obtained by multiplying story points with impact factors and summing the results. The *CPV* is the fusion of predictive variables that prognostically measures the target variable(s). It has also been realized that the effects of degradations or improvements are accumulated over releases creating the need for a cumulative damage model [30]. This methodology assumes that all the degradations incurred in each release accumulate until some external source actively repairs the system. Hence, the *CPV* is cumulative. Also, there are fixes/enhancements, such as improvements in processor speed or removal of features that demanded lots of processing space causing the application to run slow, may result in improvement of software performance. To realize the benefits of such events, the story points of such faults and enhancements are labelled as negative. A negative story point reduces the cumulative *CPV* and can even result in a negative *CPV*, which, subsequently, estimates improvements in software performance.

In the next step, natural language processing (NLP) is used for the automated classification and estimation of story points since faults and enhancement requests come in descriptive (text) format. A Naïve Bayes Classifier is used for such classifications and estimations. Naïve Bayes is a popular algorithm for text classification as it is highly efficient, easy to use, requires a small amount of training data, and less pre-processing of data. It, also, performs well in multi-class prediction [31]. Naïve Bayes classifies the faults and enhancement requests into multiple classes of pre-defined story points (Fibonacci sequences). To prepare data for NLP we removed punctuation and stop words, performed sentence segmentation, word tokenization, predicted parts of speech, performed text lemmatization, dependency parsing, found noun phrases and performed name entity recognition. The performance of Naïve Bayes is verified and validated by computing a confusion matrix and comparing actual (test data) results with classified results. The posterior probability of class (*c*, target) given predictor (*x*, attributes) is given by,

$$P(c|x) = \frac{P(x|c)P(c)}{P(x)} \tag{2}$$

where *P(c)* is the prior probability of class, *P(x|c)* is the likelihood, which is the probability of predictor given class, *P(x)* is the prior probability of predictor.

In the next step, clusters of analogous releases are formed by k-means clustering. The *CPV* is used to identify the number of clusters. Cleaned text data (faults/enhancement requests) is transformed into Unicode before plugging it into k-means to form clusters. Every time a new data point(s) is available, clustering is carried out to identify appropriate cluster(s) for the new data and to verify if any new cluster(s) has emerged. The objective function to identify the clusters (*k*) of analogous releases is given by,

$$J = \sum_{j=1}^{k} \sum_{i=1}^{n} \left\| x_i^{(j)} - c_j \right\|^2 \tag{3}$$

where *k* is the number of clusters, *n* is the number of cases, $x_i$ is case *i* and $c_j$ is the centroid cluster *j*.

The clustering process does not contain a ground truth label, which makes its validation tricky. However, the number of clusters is identified and validated based on an elbow diagram (within-cluster Sum of Square). In the elbow method the explained variation is plotted as a function of the number of clusters and the elbow of the curve is chosen as the number of clusters to use.

### 3.4 Prognostic Model

There are several paradigms for solving PHM problems, including, data-driven and model-based approaches. Model-driven approaches start with an understanding of how a physical system works and how it can fail. Whereas data-driven approaches are enabled by algorithms determined from machine learning without focusing on the underlying processes [32]. Implementation of model-driven approaches requires a deep understanding of the complexity of the system. In the case of software systems, complexity includes the specific set of characteristics of the code and its interactions with other pieces of code, with device drivers and with operating systems. An accurate understanding of the complexity may not be available for the software system, especially for a new software system. Software systems can also be very versatile and can change frequently. However, for software systems, controllable test beds (staging environments) can be generated, and appropriate data can be populated. The availability of data, even for a new software system, is not a big challenge, but still models are needed to predict and get the meaning of data sets. Therefore, a fusion approach, which combines a data-driven approach with a model-driven approach, has been used for this methodology. The configuration of earlier versions of the software system can be accomplished in controlled test beds if the corresponding artifacts are available, or if historical data are available. Control over the environmental variables is possible in such test beds. Due to advancements in sensor technology, computational capabilities, and dedicated software/hardware interfaces, fusion approach can carry out highly accurate fault detection when the system monitoring data for nominal and degraded conditions are available [33].

Based on the estimated *CPV* and historical performance data, performance parameters for the future releases are predicted using appropriate prognostic model(s). Given the problem type (supervised) and the properties of the predictive and target variables (structured), prognostic model(s) selection is motivated by data volume, relationship between variables (linear/non-linear/correlated/noncorrelated) and the number of variables. The RUL is identified by plotting both the historical and predicted performance parameters against the usage parameter and by defining a threshold. As discussed earlier, the usage parameter is *software releases*. There are no industry standards for performance thresholds. Instead, thresholds are defined and agreed upon by the end users and vary from software system to system and from end user to end user. RUL changes continuously as the number of faults and future enhancement requests change. This makes this methodology dynamic. This methodology also provides options to control the contents of future releases.

## 4. Methodology Demonstration

For demonstration, Bugzilla has been chosen as a case study. Bugzilla is an open-source tool that is used to track defects in software applications.[5] End users and the development team report faults and enhancement requests on Bugzilla in an open-source repository (https://bugzilla.mozilla.org/).

---

[5] The demonstration in this paper is on the Bugzilla application itself, not to be confused with the function of Bugzilla, which is to capture and track defects in other software applications. In this demonstration we are capturing and tracking faults and enhancements in the Bugzilla application itself.

Some of these requests have been addressed and the rest are work in progress. For demonstration purposes we focused only on response time (RT) as a performance parameter (target variable) and faults and enhancements as predictors that are related to RT only. However, the methodology described in Section 3 can be applied to other performance parameters as well. Test beds were built within standardized, scaled and controlled, staging environments on Microsoft Virtual Machines. Response time (RT) data has been populated by preparing multiple separate test beds corresponding to individual versions of Bugzilla. Open-source data has been gathered on faults and enhancements, reported by Bugzilla end users and developers [34]. Bugzilla is a web-based system but, for the purposes of this demonstration, it needed to be installed on local server. Using a version of the application on a local server allowed Bugzilla's performance to be observed as a function of changes made both in the software and hardware components. Bugzilla is written in Perl, which is able to run on cross-platform operating systems and is compatible with major databases, including SQL and Oracle. Bugzilla has multiple integrated modules, including, but not limited to, Perl, Apache, SQL, PostgreSQL, Oracle, and SQLite. Figure 2 shows Bugzilla's architecture. Faults and future enhancement requests are reported based on functional, regression and performance test results, end user experiences and requirements set by the end user groups. Much of this data is collected automatically, using software health monitoring tools.

Bugzilla was chosen for demonstration because it has been in existence for over 22 years and many of its previous versions are still available and configurable. Most of its historical release notes, faults and enhancement requests are available open source. The actual applications and functionality of Bugzilla are irrelevant to this demonstration.

Test beds produced historical response time (RT) over multiple releases in controlled and standard test (staging) environments. Test bed specifications are:

- Microsoft VM ware (Windows 10)
- RAM: 4 GB
- HDD: 50 GB
- Processor: 64-bit
- Available space on HDD after installing all the required software and tools 25 GB.

The installed versions (also known as releases in this case) of Bugzilla were: 3.6.3, 4.4, 4.4.9, 5.0, 5.0.1, 5.0.2, 5.0.3, 5.0.4, 5.0.5 and 5.0.6. Selection of these versions was motivated by the availability of configuration files. The selection of releases was kept as consecutive as possible to easily capture all the fault fixes and future enhancement requests implemented in each release. Though there were other releases in between release 3.6.3 and 4.4, and 4.4 and 4.4.9, due to the unavailability of all the associated functional files, no releases in between the above-mentioned releases were configured in test beds. However, all the faults and enhancements data, in between those releases were successfully collected. Faults and enhancement requests that were populated in between releases but were never fixed, were used to architect future releases.

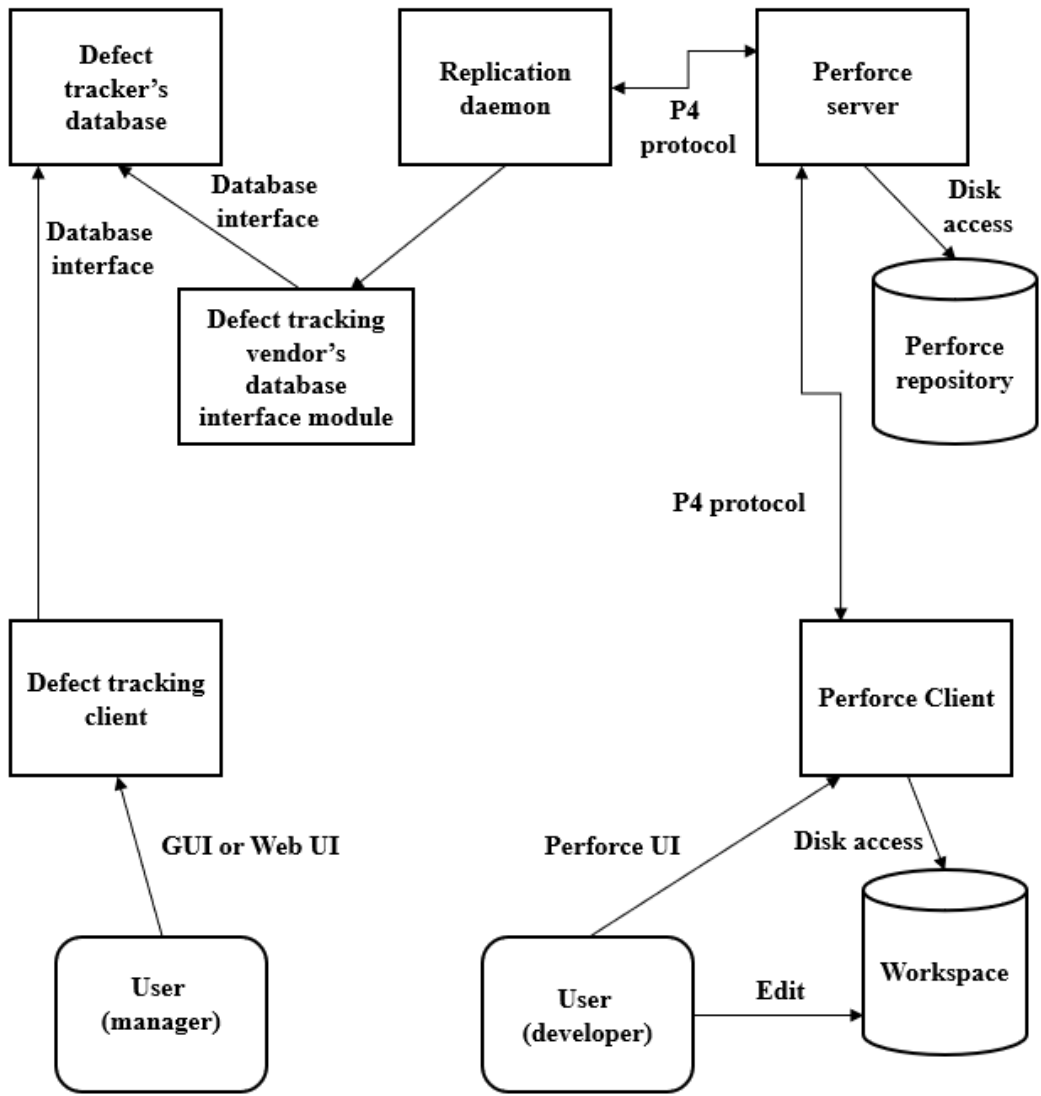

**Fig 2**. Bugzilla's architecture [13].

Software is, generally, divided in to two major segments: system software and application software. System software keeps everything working and application software allows end users to complete tasks. In this demonstration, RT was measured by varying both the application software and the system software. First, RT was measured by changing the versions but keeping all the other parameters constant across the test beds. The idea is to identify how performance parameters are affected by code changes in the application software. Then RT was measured by changing processor speed (32-bit versus 64-bit) while keeping the remaining parameters unchanged. We configured Perl 5.32.0, Java 15.0.1, SQL 5.0.15 and Apache 2.4 across all the Bugzilla versions.

The collected data types were:
- Unstructured data (texts): Faults and future enhancement requests
- Categorical variables: Fault/enhancement classes
- Continuous variables: Story points, impact factors, Cumulative Predictive Variable (SPV)

As mentioned earlier, there are no industry standards for thresholds of software performance parameters including response time (RT). Rather, threshold is determined based on de-facto standards that vary from application-to-application and governed by end user requirements. However, there are three main time limits for RT to keep in mind when optimizing web performance [35]:

- 0.1 second: The time for the end user to feel that the system is reacting instantaneously
- 1.0 second: The limit for the end user's flow of thought
- 10 second: The limit for keeping the end user's attention.

For this demonstration, a RT threshold of 9 seconds is considered.

## 4.1. RUL Calculation

JMeter [36] was used to measure RT for the ten different releases to identify the effects of version (code) changes. RT was also measured for Bugzilla versions 3.6.3, 4.4, 5.0.4, 5.0.5 and 5.0.6 to identify the effect of processor speed changes. Following are the environmental properties used in JMeter (version 5.4) to create virtual loads to populate data on RT:

- Number of threads (users): 5
- Ramp-up period (seconds): 5
- Loop Count: 5

Here, the number of threads represent the number of virtual users that will connect to the web service at once and loop count is the times individual users connect to the web service. The ramp-up period represents how long it takes to "ramp-up" to the full number of threads chosen. For each page of Bugzilla, 5x5 (number of threads) (loop count) = 25 readings were taken, and averages were calculated. After taking 5 averages in this way, an average of these 5 averages was calculated. This was carried out for all the pages of Bugzilla, including: Home, New, Browse, Search, Reports, Preferences and Administration. In the end an overall average of RT was found. The advantage of using this type of averaged response time is that it is minimally affected by fluctuations in sampling and considers all the values in the series. However, the averaging approach used is specific to this demonstration, other approaches may be appropriate for other performance parameter and other types of software systems. Also, software performances are only impacted when corresponding faults get generated or fixed. Hence, the impact of faults is to be counted only in the releases at which it was fixed or generated.

Figures 3 and 4 show the collected data with lognormal trend lines. Reduced response time means increased performance. According to these results, the response time increases as *CPV* increases. However, in some releases, RT decreased as certain improvements were made in those releases.

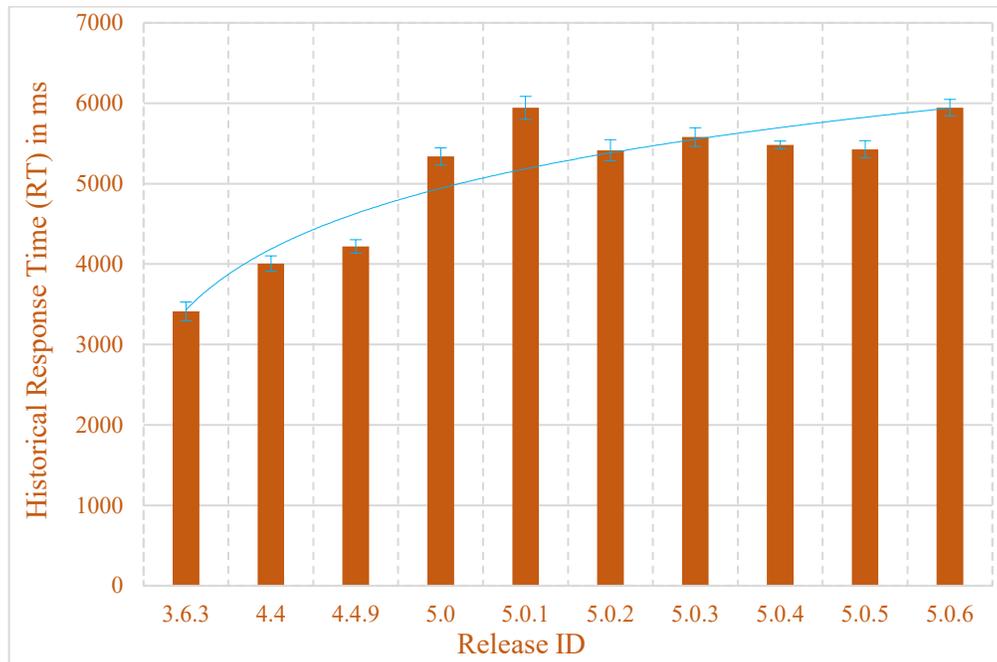

**Fig 3.** Response time (RT) over multiple Bugzilla releases.

To address such behaviors, negative *CPV*s have been accounted for. As an example, in Bugzilla version 5.0.4, a graphical report feature was removed. Elimination of this feature had a positive impact on RT, since graphical features make applications run slow. The resultant *CPV* specific to that release was negative (-0.75) and the cumulative *CPV* was 37, 0.75 down from the earlier release (5.0.3). A decrease in response time had been observed in this release, which is an improvement. Similarly, in release 5.0.5 the code was reformatted according to the same conventions as the popular Mojolicious product [37] and the bugs_fulltext table was then in InnoDB instead of MyISAM. This caused inconveniences in upgrades but resulted in improved performance. The cumulative *CPV* in this release was 36.5, a reduction (-0.5) from the previous release of 5.0.4 where cumulative *CPV* was 37.

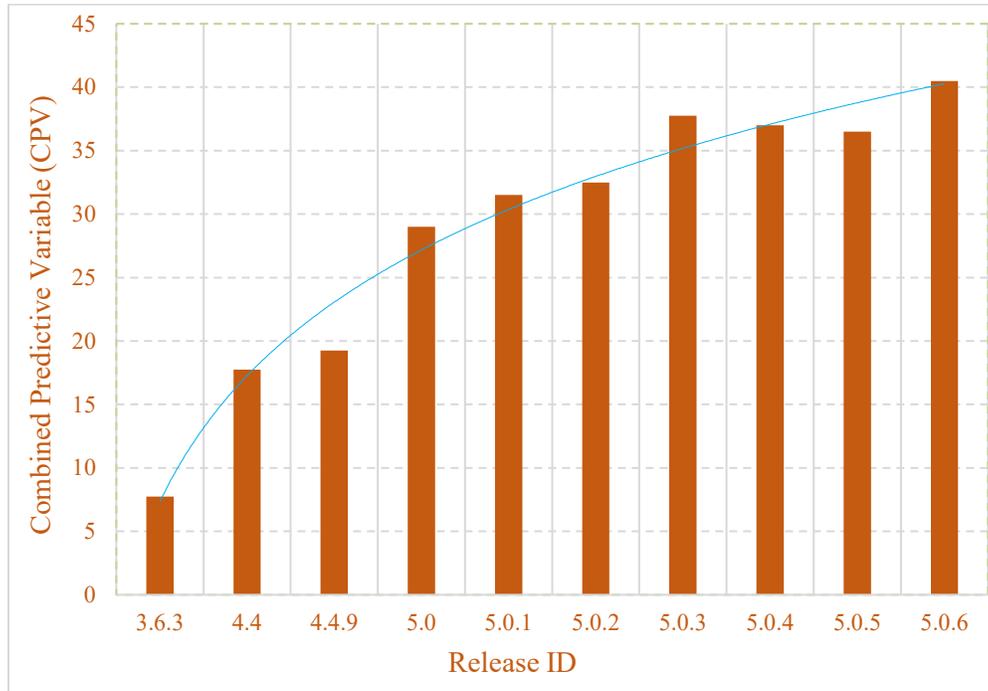

**Fig. 4.** *CPV* over multiple Bugzilla releases.

Figure 5 shows the impacts on RT due to changes in processor speed. Obviously, RT is higher in case of 32-bit compared to 64-bit in all the releases as 64-bit has a higher processing speed.

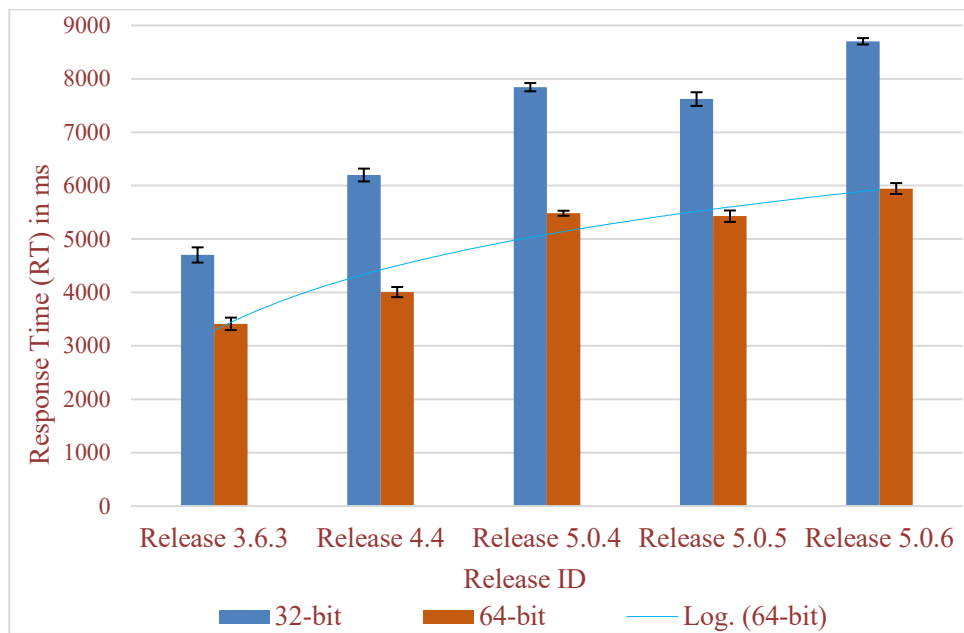

**Fig. 5**. Response time (RT) over 32-bit vs 64-bit multiple Bugzilla releases.

Next, data cleaning, classification and estimation of *CPV* clustering was carried out to identify analogous releases. To statistically validate the numbers of clusters, an elbow within-cluster Sum of Square) diagram (Figure 6) was formed. As shown in Figure 6, 3 is an appropriate number of clusters for this case as the elbow to the curve starts at 2 clusters. However, due to the limited amount of data, no $R^2$ or *p-values* could be achieved when running predictive algorithm (regression) on one of the clusters. Hence, for this demonstration we have used 2 clusters instead of 3.

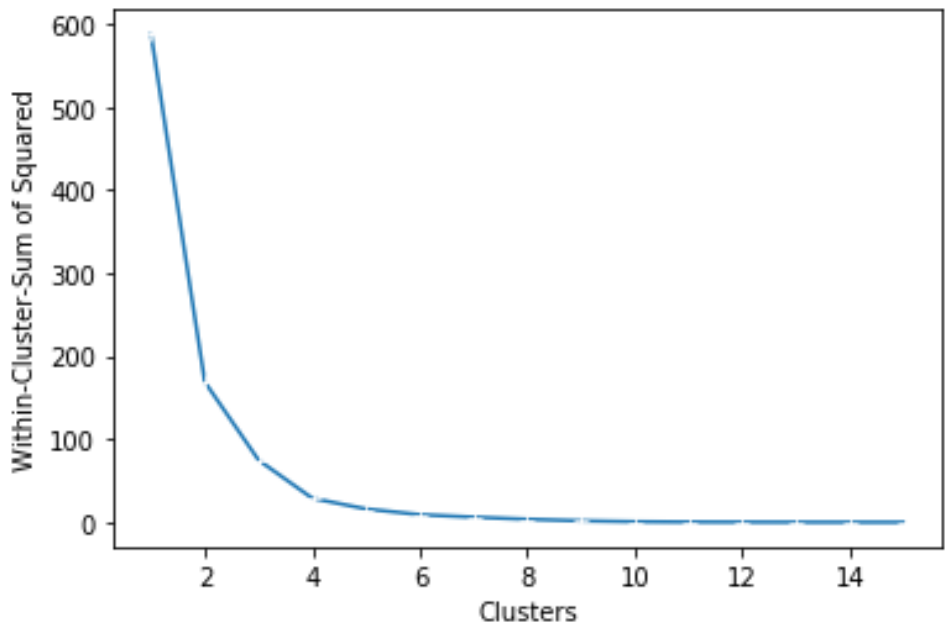

**Fig. 6.** Elbow diagram.

Google Colab Notebook was used to write and run the classification and clustering algorithm in Python. Table 3 shows the clustering results. Release 3.6.3 to 5.0.6 are *historical* releases and releases 6.0.0 to 7.5.1 are *future* releases. Among the future releases, 6.5.0 appears to be in the same cluster (B) as release 5.0.2, 5.0.5 and 5.0.6. The remainder of the releases are from the other cluster (A).

However, with the availability of more data, this relationship can be evaluated further, and new model(s) or ensemble techniques may be applied. The prognostic model was validated in several different ways. First, we compared actual data (on RT), generated by the test beds with the predicted data. Then, the model was validated statistically based on the adjusted $R^2$ value. The higher adjusted $R^2$ value along with a low *p-value* ($< 0.05$) indicates that we can reject the null hypothesis, i.e., the changes in the target variable are associated with the changes in predictive variable. Lastly, a k-fold cross validation was carried out to determine fitness of the model.

**Table 3:** Analogous releases based on k-means clustering.

| Release | Cluster |
|---------|---------|
| 3.6.3 | A |
| 4.4.0 | A |
| 4.4.9 | A |
| 5.0.0 | A |
| 5.0.1 | A |
| 5.0.2 | B |
| 5.0.3 | A |
| 5.0.4 | A |
| 5.0.5 | B |
| 5.0.6 | B |
| 6.0.0 | A |
| 6.5.0 | B |
| 6.5.1 | A |
| 7.0.0 | A |
| 7.5.0 | A |
| 7.5.1 | A |

A higher cross validation score confirms that the model is a fit for this demonstration. For validation purposes it is not necessary that the software under test reaches its end of life since historical release data can be generated and used as test data/validation data to compare with. Since RUL is estimated by plotting performance parameters (e.g., RT) against release cycles, based on a user defined threshold, validation of the RUL is dependent upon validation of RT.

R-Studio was used as the analytics platform to predict response time (RT) based on the regression analysis.

### 4.2. Case Study Results

To identify the correct prognostic model(s), a correlation analysis (Figure 7) was carried out to evaluate the relationship between the target variable (RT) and the independent variable (*CPV*). Quadrant I show a higher degree of association (0.95) between RT (Quadrant IV) and *CPV* (Quadrant II). The direction in Quadrant III confirms the positive nature of this relationship, i.e., as the *CPV* increases, RT increases. The *x* axis in Quadrant I represents RT, and the *y* axis represents *CPV*, the *x* axis in Quadrant III represents *CPV* and the *y* axis represents RT. Quadrant II and Quadrant IV represents distribution of corresponding *CPV* and RT. Given the high (0.95) correlation we applied linear regression, Eq. (3), to estimate RT based on *CPV*,

$$y = mx + b \tag{3}$$

where, *y* is the target (RT) variable, *x* is the predictor (*CPV*), *m* is the slope and *b* is the y-intercept.

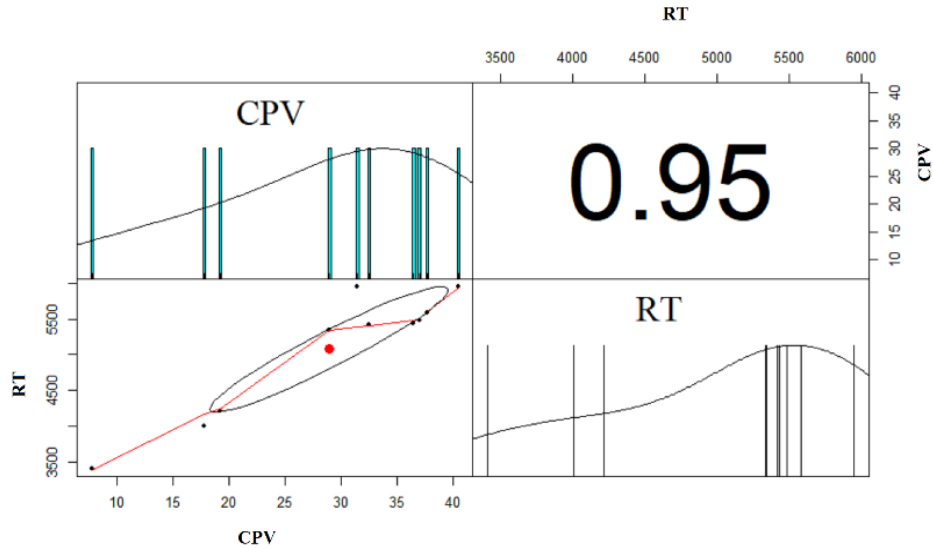

**Fig. 7**: Correlation between *CPV* and RT.

A residual analysis was carried out to validate this selection of linear regression model. The residual plot (Figure 8) shows that the data points are randomly dispersed across the horizontal axis. This confirms the linear regression model is appropriate for this data.

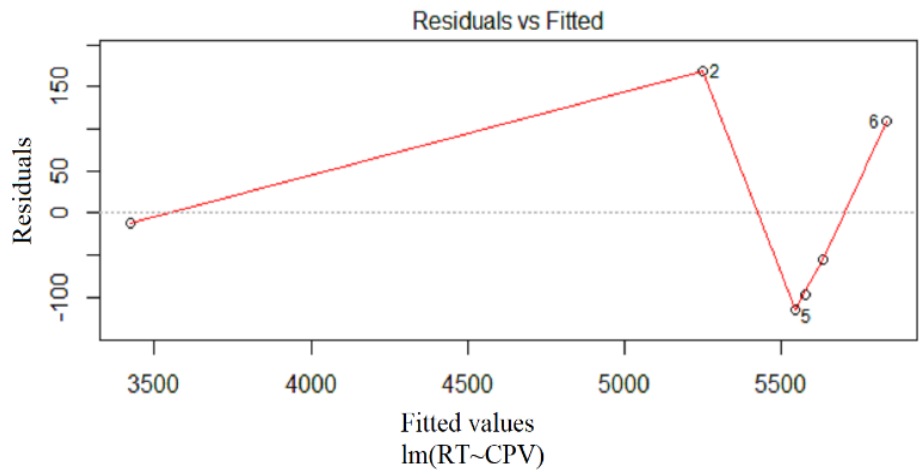

**Fig. 8.** Residual analysis plot.

The historical data was divided into training (80%) and test (20%) data sets. After training the univariate linear regression with the training data set, the performance of the model was validated using the test data set (Table 4).

**Table 4:** Actual versus predicted RT based on test data.

| Release | Cluster | Actual RT (ms) | Predicted RT (ms) |
|---------|---------|----------------|-------------------|
| 5.0.2   | B       | 5416           | 5248              |
| 5.0.3   | A       | 5579           | 5634              |

As more data becomes available other prognostic algorithms and their validation methods (e.g., cross-validation) could be explored. Table 5 shows results of the statistical validation:

**Table 5:** Model performance summary.

| Model Performance Indicators | Cluster A | Cluster B |
|------------------------------|-----------|-----------|
| Multiple $R^2$ value         | 0.9839    | 0.8968    |
| Adjusted $R^2$               | 0.9798    | 0.8796    |
| *p-value*                    | 0.00009   | 0.00036   |

For both the clusters multiple $R^2$ values and adjusted $R^2$ values are higher with significantly smaller *p-values* ($< 0.05$). A k-fold cross validation was carried out to confirm the fitness of the model. This cross validation also confirmed that the model is not overestimating or under estimating. We have used k = 2 (2-fold cross validation) due to the limited amount of data and made a 70-30 split between training and test data. The average validation score is 0.9629, which is an indication of good fitness to fit.

After training the model, future releases (six in this case) were architected for the same contents but with four optional combinations (referred to as Combos 1-4). The combos are the different combinations of the same volume of fault fixes/enhancement implementations within the same quantity of releases. Different combinations help users in planning which combo to adapt for the longest performance cycle. The release IDs are 6.0, 6.5, 6.5.1, 7.0, 7.5 and 7.5.1. The release contents are comprised of unresolved faults/expected future enhancements. However, it is possible to architect as many combinations as we like. Response time (RT) is estimated by plugging this data into the trained model and the RUL is identified (see Figure 9) by plotting them against release cycle. Based on the RUL plot it can be determined which combination of releases provides with the best RT sustainability. Sustainability refers to the combination of releases that gives us the maximum number of releases before the software system reaches its threshold, i.e., the longest RUL. In this demonstration, Combo-1 had the longest performance cycle. In Combo-1, release 6.0, several new faults emerged, e.g., JSON parsing error, query returning more comments then called for and components popping up twice unexpectedly. These faults increased the RT. Also, a few enhancements were made, e.g., user interface (UI) was enhanced, system message architectures were changed and others that increased the RT as well. Few faults were fixed, e.g., index.cgi (home) was made useful for logged-in users, all Bugzilla JavaScript was moved into separate files and others, that slightly decreased the RT. The accumulated *CPV* for release 6.0 into Combo-1 was 19. Similarly, the *CPV* for the subsequent releases were: -4.25, 4, 2,3 and 16.75. The same exercise was carried out for the three other combos. After prognostically measuring RT and plotting in the RUL graph (Figure 9) if developers plan releases according to Combo-1, they will obtain the highest number of releases before the software reaches the RT threshold.

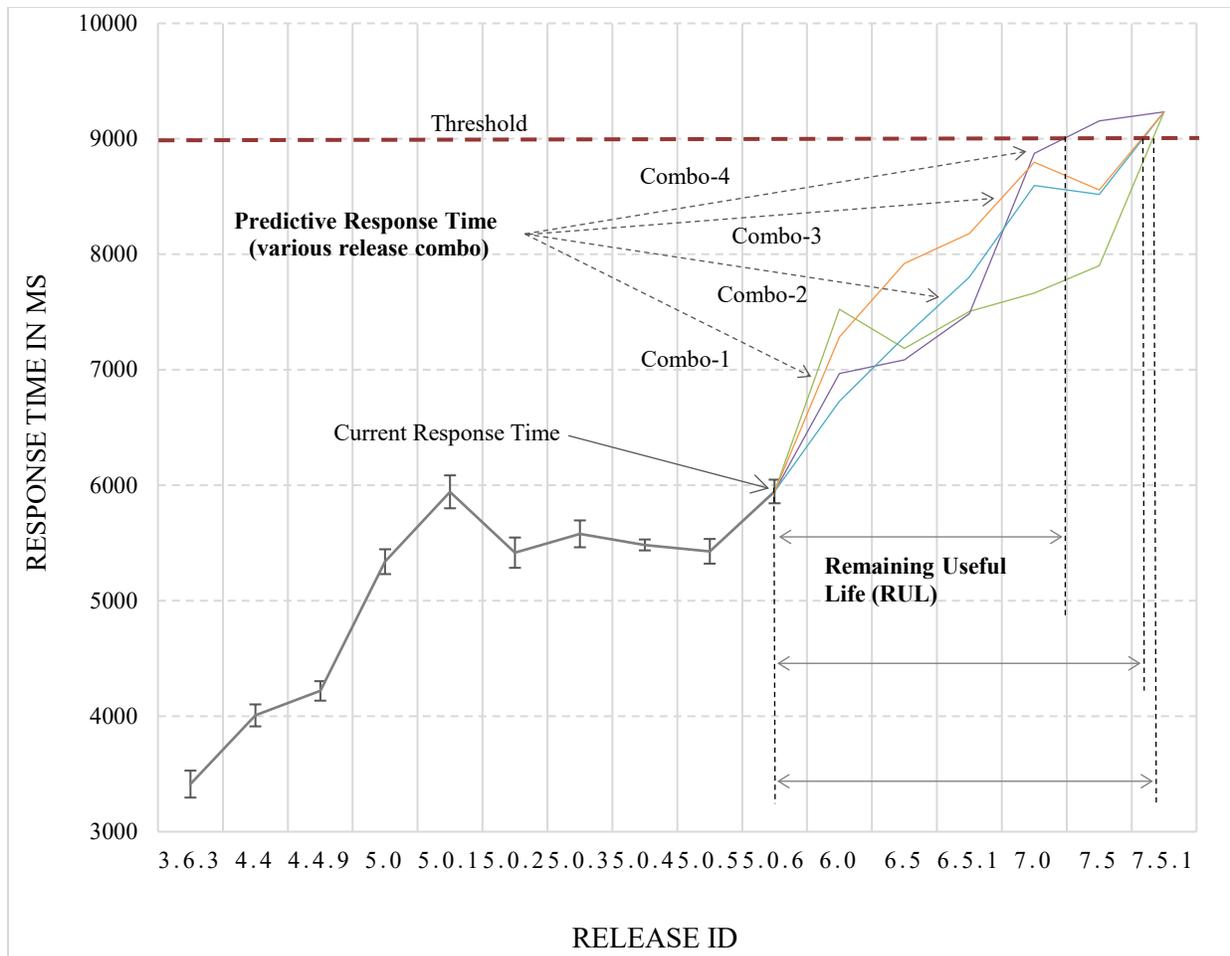

**Fig. 9.** Estimation of Remaining Useful Life (RUL) for Bugzilla.

Note, the accumulated *CPV* for release 6.5 is negative as more enhancements, that improve RT, were expected to be introduced and RT improving faults were fixed in this release. As an example, loading time of query.cgi was improved with large number of products/components, updates were made to improve compatibility with the Oracle server, system message architecture was improved, JSON parsing error was fixed, code fixes were carried out so that components do not show up twice in the component list anymore and few other fault fixes and enhancements were carried out.

As discussed earlier, this methodology provides the development team with better control over future release planning for software updates/upgrades after they identify the critical path for the longest RUL for the software system. The number of releases and the points at which the developers will need to do software update, upgrade, and others, can be predicted, and controlled.

Using the methodology described in this paper, one could find a resultant RUL by combining RULs for all the relevant performance parameters. Different performance parameters are prioritized differently for each software system and for each end user group. In the real world, end users always make trade offs among performance parameters based on their individual requirements. For example, some end users, RT could be a priority and for the others availability (even with a lower RT) could be the priority. Also, there are cases when the implementation of a single fault/enhancement request may affect multiple performance parameters. In such cases both the end users and the development team will want to visualize how any future changes could affect individual performance parameters. As an example, updating a version of Bugzilla to make it compatible with a newer version of the SQL database may result in a better (shorter) response time but could reduce the availability if there were any coding errors, or interphase errors or compatibility issues.

## 5. Discussion and Conclusion

This paper demonstrates that Prognostic Health Management (PHM) concepts and techniques can be utilized for software systems. While software maintenance can benefit from PHM, PHM also, dynamically predicts the point beyond which a software system would need to be abandoned instead of maintained. PHM can assist with relevant budgeting and scheduling as well. Although, more usage parameters need to be explored, release cycle appears to be an appropriate usage parameter for many software systems, given that software degrades/decays and performances change when upgrades and updates are made through releases. The demonstration presented addressed one performance parameter (response time) and a similar methodology could be followed if other performance metrics are to be considered. If other performance parameters are used, the corresponding input variables that affect them will need to be identified and mapped. Then the steps presented in the methodology would be followed to estimate the RUL in terms of the new performance metrics. Corresponding modifications to how data is collected, processed, and analyzed would be necessary. In future research other models will be explored and model accuracy will be compared.

To apply this methodology, predictive variables and target variables need to be identified, aligned and *CPV* needs to be estimated (note, the methods to estimate *CPV* may vary depending up on input data types). Calculating the RUL with multiple options ("combos") assists both the end users and the development team in estimating number of releases that they can plan before performance thresholds are reached. At the same time, it provides them with control over the quantity and the contents of releases by planning the *CPV* accordingly. The RUL of the software system provides the development team with the ability to visualize the past and future releases in graphs, which assists in planning and decision-making.

Other benefits involve the mitigation of equipment failure. Timely identification of RUL trends allows developers to anticipate and avoid issues that could limit software performance and or reduce software system life. When a systematic cause is observed, developers can adjust their responses accordingly to minimize problems. Also, over time, the developer's knowledge of the software system(s) is enriched by continuously developing databases of release details. Such knowledge, gathered over time, can enhance the accuracy of future diagnosis and prognosis [38].

The methodology described in this paper is potentially applicable to all software types that have performance requirements (including: application software, system software, programming software and driver software). As an example, even real-time programs need to guarantee *response* within specified *time constraints* that are, often, referred to as "deadlines" [39]. Such responses are often understood to be on the order of milliseconds, and sometimes microseconds. As an example, a pacemaker is a real-time system that needs to operate fast enough to guarantee the health and safety of the patient [40]. RT is one of the most important performance criteria for any real-time system. In the similar way, embedded software is designed to meet real-time system requirements. Such software requires precise real-time responses to the microsecond in a distribution system and the system is expected to be fault tolerant under strict timing requirements [41].

The proposed methodology is a fusion and considers the descriptions, frequencies and impacts of faults/enhancement requests and is independent of the inherent software mechanism and software architecture. Therefore, this methodology is independent of types of coding language(s), numbers and types of integrated components, and types of associated hardware. To apply the proposed fusion methodology, the development team is not required to have expert knowledge on inherent mechanism of any software system, though that would be beneficial. However, variations among *CPV* and estimations of output parameters even for the same software system on different environments are expected. As an example, this methodology will work equally well for a software application that is hosted either on local machines (e.g., local server-based software system) or on cloud environments. This is, because, in both the environments software system will have faults and future enhancement requests. Even though their numbers and estimated *CPV* may vary, as long as there are data on faults and enhancements, the discussed methodology will be applicable. The same is true when it comes to changing processor speed.

Future work will focus on formulating a robust function that will combine the impacts of multiple influencing factors (e.g., code changes, processor speed changes) and measure a resultant RT.

Future work will also involve estimation of RUL based on other performance parameters as stated in Section 3.1. Other models, including classical models for software reliability and different classification/clustering models, will be explored as well. Keeping RUL visualization separate for each performance parameters would be better so that user can easily pick and choose to decide while making performance tradeoffs. Different software systems, like mission-critical software, big data tools (e.g., Spark, $H_2O$ and others) could be used to experiment with and to estimate RUL.